\documentclass{aa}
\usepackage{graphicx}
\begin{document}

\title{On the relation between the IR continuum and the 
active galactic nucleus in Seyfert galaxies\thanks{Based on observations
with ISO, an ESA project with instruments funded by ESA
Member States (especially the PI countries: France, Germany, the Netherlands
and the United Kingdom) and with the participation of ISAS and
NASA.
}}
\author{M. Almudena Prieto\inst{1,} \inst{2}
A. M. P\'erez Garc\'{\i}a\inst{2} and
J. M. Rodr\'{\i}guez Espinosa\inst{2}}
\offprints{A. M. P\'erez Garc\'{\i}a}
\institute{European Southern Observatory, D-85748 Garching, Germany
\and
Instituto de Astrof\'\i{}sica de Canarias, E-38200
La Laguna (Tenerife), Spain }


\date{Received         / Accepted }

\abstract{A sample of the brightest known Seyfert galaxies  
from the CfA sample 
is analyzed on the basis of ISO photometric and spectroscopic data.
 Regardless of the Seyfert type, the  mid-IR  continuum emission  from
these galaxies  is found to
be correlated with the coronal line emission arising in the nuclear
active region. Conversely,  the correlation degrades progressively 
 when moving from the mid- to the far-IR emission, where it
ends to vanish.
It is  concluded that the mid-IR emission is largely
dominated by dust  heated by processes associated with the active
nucleus whereas the far-IR
is a different component most probably unrelated with the active region.
We suggest that the far-IR component is due to  
dust heated by the stellar population in the disks of these galaxies.
\keywords{galaxies: Seyfert--infrared: galaxies}
}
\titlerunning{On the relation between...}
\authorrunning{Prieto et al.}
\maketitle

\section{Introduction}

Dust reprocessing of some primary emission is at present regarded as
the most likely source for the mid- and far-IR emission in Seyfert
galaxies.  Whether dust is predominantly heated by 
starlight, processes associated with the central power source -
photoionization, shocks -  or a mix of both may depend on the
region of the IR spectrum under study.
The mid-IR emission was  found  by Edelson et al. (1987)
to be extended on scales of about 1 kpc in Seyfert 2s and
many Seyfert 1s, and therefore most probably related to dust heated by 
star formation regions. This result has been 
corroborated  by several authors (Giuricin et al. 
1995; Maiolino et al. 1995) using much  larger samples of Seyfert galaxies.
Regarding the far-IR, Rodriguez-Espinosa et al. 
(1986)  concluded that a large fraction of the IRAS far-IR emission
in Seyfert's is due to dust heated by circumnuclear starforming events.
This is also
confirmed by Giuricin et al (1995) and Rigopoulou et al.
(1997)  among others mainly on the basis of IRAS data.
More recently, P\'erez Garc\'{\i}a et al (1998) and 
P\'erez Garc\'{\i}a \& Rodr\'{\i}guez Espinosa
(2001) have shown on the basis of ISOPHOT data that the spectral energy
distribution out to 200$\mu m$ of  Seyfert galaxies  
is of thermal origin and the
sum of three different components: 
a warm component peaking at about 16~$\mu$m,
 related with the circumnuclear and nuclear emission; 
a cold component peaking  at
$\sim$60$\mu m$, produced by dust heated by starforming regions in the
host galaxy; and a very cold component peaking  at $\sim$150$\mu m$, 
emitted by dust heated by the general
interstellar radiation field. These authors further show that the
far-IR emission in those objects for which  ISOPHOT 90 $\mu m$
maps are available, can be traced out to an extension which is similar
or larger than 
that seen in optical R band images (P\'erez Garc\'{\i}a et al 2000).

In this work, we would like to further examine the nature of the
aforementioned  IR components,  in particular, in regard to 
the contribution of the central AGN
source. In doing so, the ISO continuum emission of a sample of very
bright Seyfert galaxies is  compared with a reliable tracer of
the AGN activity, that is their  coronal line spectrum.  Coronal lines
are unique tracers of the AGN activity, the reason being the high
ionization potential of the species that give origin to them. 
These lines form in very energetic
environments and close to the excitation source, thus tracing 
  a rather  compact region close to the central AGN. 
The coronal lines used in this work
require photon energies in the 50 - 300 eV range and thus their
origin is directly linked to processes associated with the AGN.  
In principle, the main dust heater in the nuclear region are both
circumnuclear star forming regions and the active nucleus. Clues on
which of those mechanisms  is predominant, and in turn to probe into the AGN
energetics may be found by comparing the coronal line spectra with the
IR continuum emission.  The relationship between
these two independent energy nuclear tracers is thus investigated herein.

\section{The Seyfert sample used in this work}

The sample of galaxies used in this work is extracted from the CfA
  sample of Seyfert galaxies. The complete CfA sample was observed with
  ISOPHOT in the wavelength range between 16 and 200~$\mu m$. The
  brightest members were the subject of a complementary spectroscopic
  study with the ISO short wavelength spectrometer ISOSWS.  This
  spectroscopic sample includes all the CfA Seyferts detected with IRAS
  at 12 $\mu m$ and is the one used in this work. It is further
 complemented with  the Southern  Circinus galaxy which
shows one the brightest ISO coronal spectra known (Moorwood et al. 1996). 
 For all these
  galaxies, observations of the coronal 
lines [O IV] 25.9 $\mu m$, [Ne V] 14.3 $\mu m$, 
[Mg VIII] 3.02 $\mu m$ and [Si IX] 2.58 $\mu m$
  were obtained.  
The reader is referred to P\'erez Garc\'{\i}a et al
  (1998) and to
  Prieto \& Viegas (2000) for further details regarding the
  observations and analysis of the ISOPHOT and ISOSWS data
  respectively.  

In total, the sample contains five Seyfert type 1 and seven type 2.
 All the galaxies present strong [O IV]$\lambda 25.9\mu m$ and
[Ne V]$\lambda 14.3\mu m$ lines whereas [Mg VIII]$\lambda 3.02\mu m$ and
 [Si IX]$\lambda 2.58\mu m$ are systematically the weakest lines
 (cf. Table 1 in Prieto \& Viegas 1999). Clear detections for the
 later are only found in Circinus, NGC~1068, NGC~4151, and possibly in
 Mrk 817. Due to the weakness of the [Mg VIII]$\lambda 3.02\mu m$ and
 [Si IX]$\lambda 2.58\mu m$ lines, the present analysis is based on the
 [O IV]$\lambda 25.9\mu m$ and [Ne V]$\lambda 14.3\mu m$ data only. The
 fluxes for these lines correspond to integrated values within the
 ISO-SWS aperture of 20x33 arcsecs. Details of these observations can
 be found in Prieto \& Viegas (2000).

The ISOPHOT data covers the 16 - 200~$\mu m$ range. Continuum fluxes
at 16, 25, 60, 90, 120, 135, 180 and 200 $\mu m$ were measured for all
the galaxies in this sample. In addition, ground based data at 10$\mu
m$ taken from Contini et al. 1999, Edelson 1978, Rieke \& Lebovsky
1978, Edelson et al. 1987, Maiolino et al. 1995 and Maiolino et
al. 1998 are used. These correspond to integrated fluxes within an
aperture between 5 and 8.5 arcsec. The 16 and 25$\mu m$ values
correspond to integrated fluxes within a 120 or 180 arcsec aperture,
depending on the size of the objects. The 60 and 90$\mu m$ data were
acquired with the C100 detector consisting of a 3x3 pixel array, each
pixel projecting on 45 arcsec on the sky. The longer wavelengths
observations were done with a 2x2 pixel array (C200), each pixel
projecting on 89.9 arcsec on the sky. Although apertures for the mid-
and far- IR fluxes are different, in all cases for each galaxy the
integrated fluxes include the entire object.  Details of these
observations are in P\'erez Garc\'{\i}a et al (1998).

All the  ISO fluxes  used in this work are well above the 
detection limit of our observations with ISOPHOT and ISOSWS.
Our detection limit with ISOPHOT is about  50 mJy at 16 and 25$\mu m$.  
This is estimated
on the basis of the ISOPHOT observations of the complete CfA sample of
galaxies.
At 60 and 90 $\mu m$, the present sample is  far above  the ISO detection
limit: the minimum flux used in this work is above 200 mJy.
In the case of the ISOSWS, our detection limit is about 
$10^{-21} W~cm^{-2}$ in the spectral regions around the [OIV] and [NeV]
lines. This  is consistent
with the minimum  line fluxes detected with good confidence by Genzel
et al. (1998) of a few  $10^{-21} W~cm^{-2}$.

\begin{figure*}
\includegraphics[width=\textwidth]{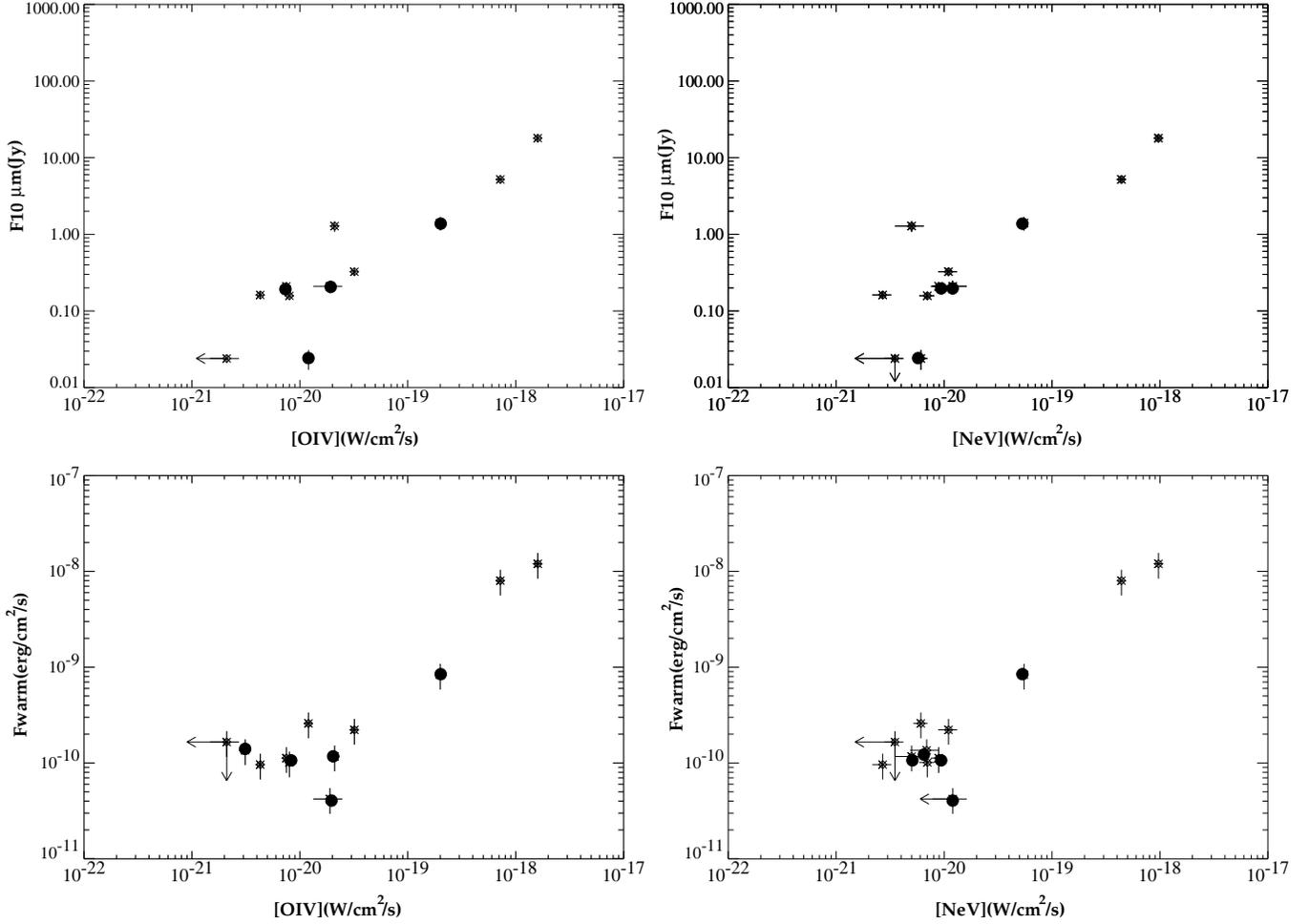}
\caption{\label{fig1}The 10~$\mu$m
continuum flux and the so-called IR warm flux as derived 
by Perez-Garcia et al (2000)
 are respectively compared with
the [O IV]$\lambda 26.9\mu$m  and 
[Ne V]$\lambda 14.3\mu$m line fluxes for the sample of galaxies
analyzed. Errors bars for the continuum and line fluxes are plotted. 
In some cases the errors are smaller than the symbols used, e.g., the 10um fluxes. Seyfert 1 objects are indicated with filled symbols, all the others 
 are Seyfert 2's .}

\end{figure*}

\section{Results}

Figs. 1 and 2 show various comparisons  between 
the [O IV]$\lambda 25.9\mu m$ and [Ne V]$\lambda  14.3\mu m$ coronal lines
fluxes and  the 10, 16, 25, 60 and 90~$\mu m$ continuum
fluxes for all the galaxies  in the sample. 
In all the figures, the brightest objects in the diagrams are  by
decreasing flux level  NGC 1068,
Circiunus and NGC 4151.

Focusing first on the mid-IR continuum emission, a  correlation is
clearly seen between the coronal line emission and the mid IR 10$\mu m$
 (Fig.1, top panels),
 16 and 25~$\mu m$ continuum fluxes
(Fig. 2 top two panels).  The linear correlation coefficient is r=0.98 for the
comparison of the 10~$\mu m$ flux with both the [NeV] and [OIV] lines,
implying a probability of chance occurrence in a uncorrelated parent
population of less than 0.1\%. The significance of the correlations decreases
as the comparison involves  longer wavelength continuum fluxes. This
may be  partially due to the fact that the 10~$\mu m$ emission is collected
from a much small aperture than the ISO apertures and hence, it should be 
less polluted by circumnuclear star formation. Formally, the 
comparison with  the 16~$\mu m$ flux yields  linear correlation coefficients
of r=0.86 and r=0.92 for the  [Ne V] and the [OIV] lines respectively (0.1\%
of chance occurrence). In general, the  correlation with 
[O IV] 25.9~$\mu m$ is better
 as this line is the strongest and  best measured in the ISOSWS
spectra (Prieto \& Viegas 2000). 

The comparison with the 25~$\mu m$ continuum flux shows a correlation 
slightly weaker than that seen with the shorter wavelengths  IR fluxes.
The correlation coefficients are r=0.81
and r=0.87 for the [Ne V] and the [OIV] lines respectively, with a
significance level of 99\%.

\begin{figure*}
\centering
\includegraphics[width=16cm]{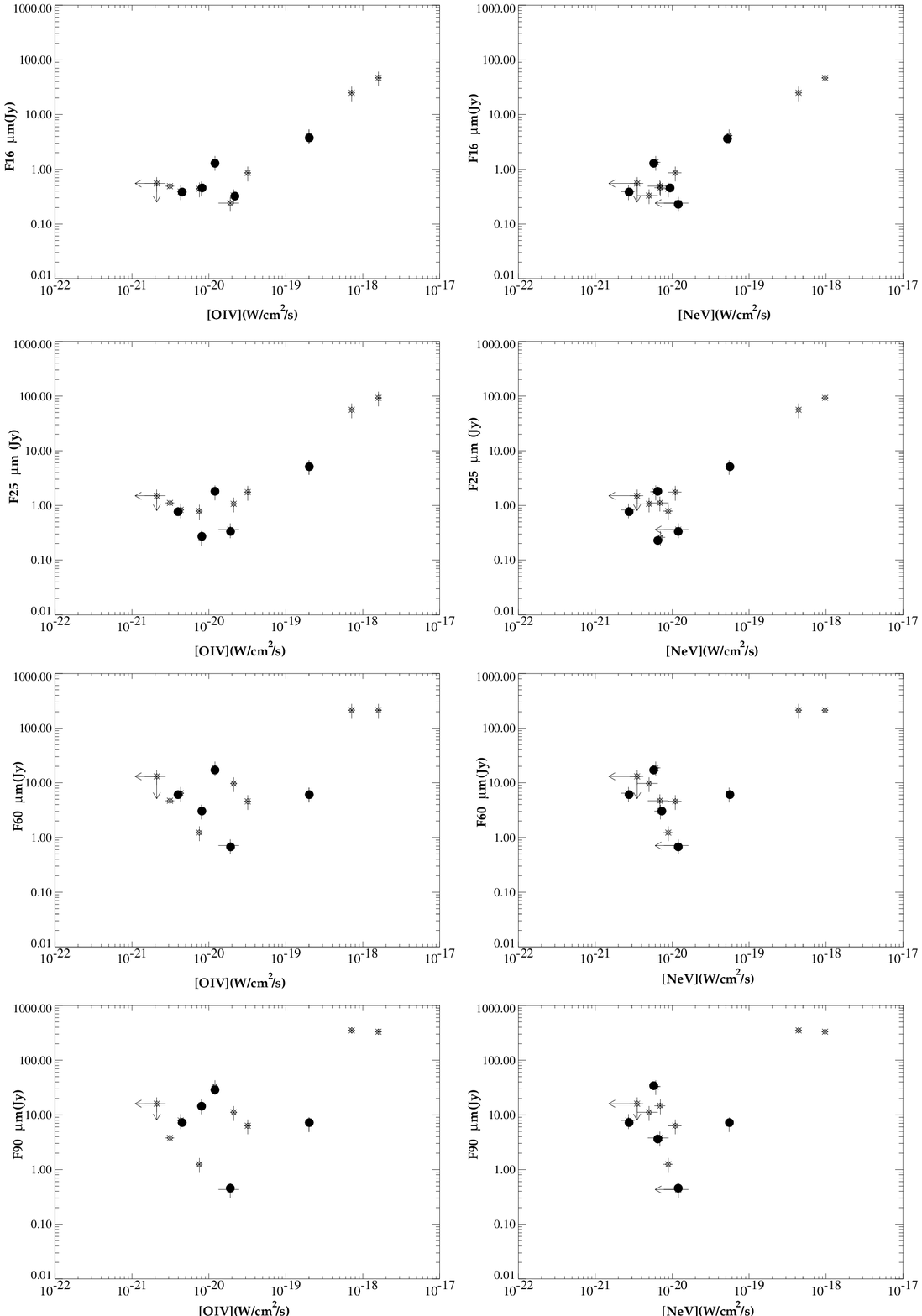}
\caption{\label{fig2}The same as in Fig. 1 but involving now the comparisons with the 
ISO 16, 25,  60 and 90~$\mu$m fluxes.}
\end{figure*}

\begin{figure*}
\includegraphics{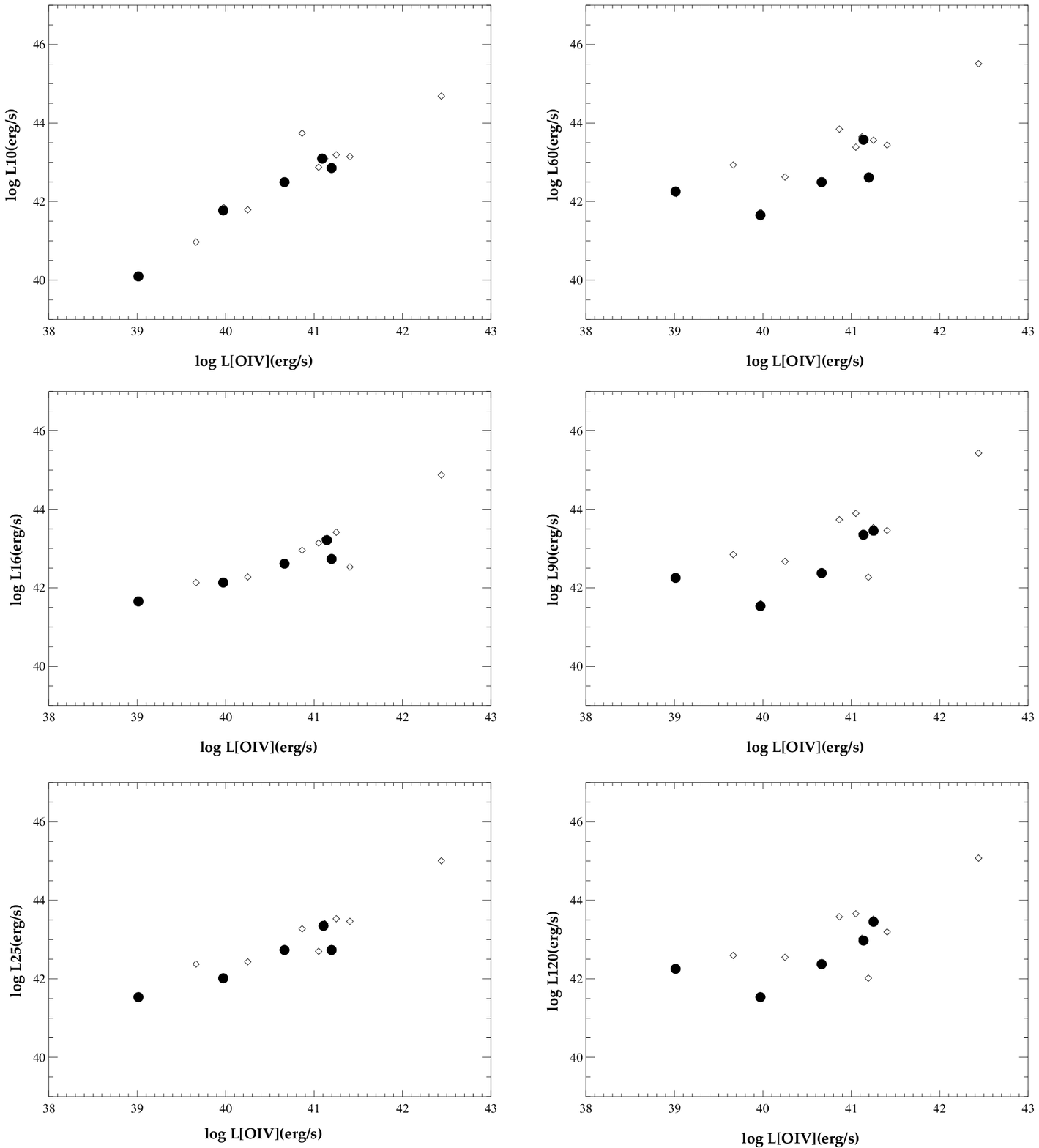}
\caption{\label{fig3} The 10, 16, 25, 60, 90 and 120~$\mu$m
luminosities
versus the [OIV]$\lambda 25.9\mu$m luminosity  for all  the galaxies
analyzed. Symbols are as in Fig. 1
}
\end{figure*}

When moving to  much cooler continuum fluxes, i.e.,
beyond 60~$\mu$m (Fig. 2 bottom two panels), the distribution of points 
becomes  rather scattered 
-- note that all panels in Fig. 1 and 2 have the same
scale. The  scatter 
increases with increasing IR  wavelength, this trend holding up
 to  the longest IR wavelength  traced with ISO, about
200~$\mu$m (note that to avoid redundancy,  
only the comparisons with the fluxes at 60 and 90$\mu$m are shown).
The correlation coefficients are accordingly lower than those derived with  
 the short wavelength IR fluxes.
For example, the  linear correlation coefficient
between the 60~$\mu$m emission  and the [OIV] line is 0.7 and between the
 90~$\mu$m 
and the [OIV] line is 0.6 -- 98\% significance
level. This result is 
compatible  with a progressive degradation of the coronal line - IR flux  
correlations as we move towards longer IR wavelengths.

The analysis of the IR Spectral Energy Distribution of the CfA  sample
of Seyfert galaxies by
P\'erez Garc\'{\i}a \& Rodr\'{\i}guez Espinosa 
(2001; cf. section 1, this work) shows the
existence of a separate warm dust component emission peaking at about 16$\mu
m$, which it is argued to be associated with the nuclear
emission. Considering  the above correlations, a tight relation
between the flux from that warm component  and the coronal emission should  also be
expected. This is shown in Fig. 1 (bottom panels).  The linear
correlation coefficients in this case are r=0.85 and r=0.90 for the
comparison with  the [Ne V] and the [OIV] line fluxes respectively,
the significance level being larger than 99.9\%.

Finally, the re-distribution of the objects in the above diagrams
when the factor distance  is considered is illustrated in  Fig.
\ref{fig3}. This figure  shows a selection of the previous
 diagrams in luminosity plots, namely the comparison  between the IR continuum luminosities -- from 10 to
120~$\mu$m--
and  the [OIV]$\lambda 25.9\mu$m line luminosity  for all  the galaxies
analyzed. In this case,  the
 brightest objects in the sample, namely NGC
1068, Circinus and N4151, show luminosities comparable to the average 
values in the sample. NGC1068 shows however the largest fluxes and luminosities
in all cases.
Regarding the distribution of objects according to their  Seyfert type,  
the five Seyfert type 1 galaxies  in the sample (NGC~4151, NGC~5548, Mrk
335, NGC5033 and Mrk817) show luminosities in both the continuum
 and the line  comparable
to those measured in the Seyfert 2 group. In all the comparisons (Fig 1 to 3),
the two groups are well mixed and follow the same overall
trend.

Fig. \ref{fig3} shows the same type of correlations as 
illustrated in the flux diagrams.
The corresponding linear correlation coefficients for the 
comparisons  involving both [OIV] and [NeV] are provided in Table 1.
As expected, the strongest 
correlation appears between the 10~$\mu$m and the coronal line luminosities.
The   progressive degradation of the correlations 
with increasing IR wavelength is now very clearly seen in luminosity plots
 (Fig. 3 
and Table 1). Note for instance the jump in correlation coefficient between
the 25 and the 60~$\mu$m quantities. 

\begin{table}
\centering
\caption{Linear correlation coefficients for the comparisons between the
IR continuum luminosities and the coronal line luminosities }
\begin{tabular}{ccc}
\hline
 & {\bf L[OIV]} & {\bf L[NeV]}\\
\hline
\hline
{\bf L10}& 0.96 & 0.95 \\
{\bf L16}& 0.85 & 0.90 \\
{\bf L25}& 0.92 & 0.87 \\
{\bf L60}& 0.80 & 0.74 \\
{\bf L90}& 0.75  &0.69 \\
{\bf L120}& 0.73& 0.64 \\
\hline
\end{tabular}
\end{table}

\section{Discussion}

Perez Garc\'{\i}a \&
Rodr\'{\i}guez Espinosa (2001) showed that the mid and far IR
emission in Seyfert galaxies arise in two different galactic environments, 
namely, the
nuclear region (circumnuclear star forming regions and the active
nucleus) and the disk of the galaxy respectively. The present results
add further support to this scenario. 

Because the coronal lines are directly linked to the nuclear activity,
the very good correlation found between the mid-IR continuum and the coronal
line fluxes
indicates that the mid IR continuum emission in
Seyfert galaxies,  regardless of their type, is largely dominated by the
active nucleus. Some scatter is present in the diagrams but this is expected 
to be  largely due to emission from circumnuclear star forming regions, which
contribution would be included in the large ISO apertures.  In
particular, the significantly better correlation with the small aperture 
10$\mu m$
flux might certainly be a direct consequence of the smaller aperture
size used in the ground based observations.

Regarding the far IR emission,  it is seen that the  
correlation between the coronal line emission and the mid and far IR fluxes
progressively
gets poorer  beyond 25~$\mu$m and longwards.  
This trend  calls   
for a different origin for the colder component, most probably unrelated with
the nuclear activity.

The most plausible source for the emission in the aforementioned IR
 components is
dust; in the case of the mid-IR, or the so-called warm component, the
correlation with the coronal line emission indicates that gas and dust
are well mixed in the nuclear environment and very likely have 
comparable efficiencies in reprocessing the primary radiation emerging
from near the central engine (see also Rudy 1984).

In the case of the far-IR, or the so-called cold component, the lack of
correlation with the coronal line emission points to a different
origin for the emission in that region.  Perez Garc\'{\i}a et al. 
 (2000) show that the 90$\mu$m
emission from the four Seyfert galaxies analyzed in their work
(NGC~5033 is the only source in common with our sample) is extended, with 
sizes comparable to those seen in optical images.
In that case, the most probable source of the emission is
dust heated by the stellar population in the disks of these
galaxies. If that is the general case for Seyfert galaxies as a class,
no obvious relation  between the far-IR continuum
and the coronal line emission should be expected, in agreement with the result
found in this work.


\begin{thebibliography}{}




\bibitem{}  Contini, M., Prieto, M.A., \& Viegas, S.M. 1999, ApJ 505, 621


\bibitem{} Edelson, R.A., Malkan, M.A. \& Rieke, G.H. 1987, ApJ 321,233

\bibitem{} Edelson, R.A. 1978, ApJ 226, 550

\bibitem{} Genzel, R. et al.  1998, ApJ 498, 579

\bibitem{} Giuricin, G., Mardirossian, F., Mezzetti, M., 1995, ApJ 446, 550

\bibitem{} Krabbe,A. et al., 1997, ApJ 476,98

\bibitem{} Maiolino, R., Ruiz,  M., Rieke, G. H. \& Keller,
L. D. 1995, ApJ 446, 561

\bibitem{} Maiolino, R., Krabbe, A., Thatte, N., Genzel, R., 1998, ApJ 493,
650
\bibitem{} Moorwood A. et al. 1996, A\&A 315, L109

\bibitem{} P\'erez Garc\'{\i}a,, A.M., Rodr\'{\i}guez Espinosa, J.M., 2001,
ApJ, in press

\bibitem{} P\'erez Garc\'{\i}a, A.M., Rodr\'{\i}guez Espinosa, J.M.,
Fuensalida, J.J., 2000, ApJ, 529, 875

\bibitem{} P\'erez Garc\'{\i}a, A.M. Rodr\'{\i}guez Espinosa, J.M., Santolaya
Rey, 1998, ApJ 500, 685

\bibitem{} Prieto, M.A., Viegas, S.M., 2000, ApJ 252, 238

\bibitem{} Rieke, G.H., Lebofsky, M.J., 1978, ApJ, 222, 49
 
\bibitem{} Rigopoulou, D., Papadakis, L., Lawrence, A., Ward, M., 1997,  A\&A
327, 493

\bibitem{} Rodr\'{\i}guez Espinosa, J.M.\& P\'erez Garc\'{\i}a, A.M.,
1997, ApJ, 487, L33

\bibitem{} Rodr\'{\i}guez Espinosa, J.M., Rudy, R.J., Jones, B., 1986, ApJ,
309, 76

\bibitem{} Rudy, R.J., 1984, ApJ, 284, 33


\end{thebibliography}
\end{document}